\documentclass[letterpaper, 10 pt, conference]{ieeeconf}

\IEEEoverridecommandlockouts                              
\usepackage{bbold}
\usepackage{hyperref}
\usepackage{amsfonts}
\usepackage{amsmath,amssymb}
\usepackage{cite}
\usepackage{graphicx}
\usepackage{subcaption}
\usepackage{hyperref}
\usepackage{wrapfig}
\usepackage{mathrsfs}
\usepackage{color}

\usepackage{latexsym,url}
\usepackage{amsmath,amssymb,array,graphicx,verbatim}
\usepackage{setspace}
\usepackage{amsfonts}
\usepackage{times}

\newtheorem{theorem}{Theorem}

\newtheorem{lemma}[theorem]{Lemma}

\newtheorem{proposition}[theorem]{Proposition}

\newcommand{\Real}{\mathbb{R}}
\newcommand{\Ex}{\mathbb{E}}
\newcommand{\Prob}{\mathbb{P}}
\newcommand{\base}{\mathbf{e}}
\newcommand\numberthis{\addtocounter{equation}{1}\tag{\theequation}}
\newcommand{\tr}{\top}

\newcommand{\R}{\mathbf{Cost}}

\begin{document}

\title{\LARGE \bf
Finite-time Analysis of the Distributed Detection Problem }

\author{Shahin Shahrampour, Alexander Rakhlin and Ali Jadbabaie
\thanks{This work was supported by ONR BRC
Program on Decentralized, Online Optimization.}
\thanks{Shahin Shahrampour and Ali Jadbabaie are with the Department of Electrical and Systems Engineering at the University of Pennsylvania, Philadelphia, PA 19104 USA. (e-mail: shahin@seas.upenn.edu; jadbabai@seas.upenn.edu).}
\thanks{Alexander Rakhlin is with the Department of Statistics at the University of Pennsylvania, Philadelphia, PA 19104 USA. (e-mail: rakhlin@wharton.upenn.edu).}} %

\maketitle

\begin{abstract}
This paper addresses the problem of distributed detection in fixed and switching networks. A network of agents observe partially informative signals about the {\textit{\textbf unknown}} state of the world. Hence, they collaborate with each other to identify the true state. We propose an update rule building on distributed, stochastic optimization methods. Our main focus is on the {\textit{\textbf finite-time}} analysis of the problem. For fixed networks, we bring forward the notion of Kullback-Leibler cost to measure the efficiency of the algorithm versus its centralized analog. We bound the cost in terms of the network size, spectral gap and relative entropy of agents' signal structures. We further consider the problem in random networks where the structure is realized according to a {\textit{\textbf stationary}} distribution. We then prove that the convergence is exponentially fast (with high probability), and the {\textit{\textbf non-asymptotic}} rate scales inversely in the spectral gap of the expected network.  
\end{abstract}

\section{Introduction}
Distributed detection, estimation, prediction and optimization has been an interesting subject of study in science and engineering for so many years \cite{tenney1981detection,tsitsiklis1993decentralized,borkar1982asymptotic,nedic2009distributed,kar2012distributed,eksin2012distributed}. Decentralized algorithms are ubiquitous in numerous scenarios ranging from sensor and robot to social and economic networks \cite{chamberland2003decentralized,bullo2009distributed,atanasov2014distributed,shahrampour2013online}.  
In these applications, a network of agents aim to accomplish a team task for which they only have {\it partial} knowledge. Therefore, they must communicate with each other to benefit from local observations. In fact, the global spread of information in the network is sufficient for agents to achieve the network goal. In most of these schemes, {\it consensus} protocols are employed to converge agents to a common value \cite{jadbabaie2003coordination,olfati2004consensus}. 

We would like to focus on the distributed detection problem in this work. The problem was first considered in the case
that each agent sends its private observations to a fusion center \cite{tenney1981detection,tsitsiklis1993decentralized,chamberland2003decentralized}. The fusion center then faces a classical hypothesis testing to make a decision over the value of the parameter. While collecting data is decentralized in these cases, the decision making part is done in a centralized fashion.

Distributed detection has also been considered in works where no fusion center is necessary \cite{cattivelli2011distributed,jakovetic2012distributed,bajovic2012large}. These works mostly focus on asymptotic analysis of the problem. Cattivelli et al. \cite{cattivelli2011distributed} develop a fully distributed algorithm based on the connection of Neyman-Pearson detection and minimum-variance estimation to tackle the problem. Jakoveti{\'c} et al. \cite{jakovetic2012distributed} propose a consensus+innovations algorithm for detection in the case of {\it Gaussian} observations. Their method possesses an {\it asymptotic} exponential error rate even under noisy communication of agents. In \cite{bajovic2012large}, the consensus+innovations method is extended to {\it generic} (non-Gaussian) observations over random networks.

Another social learning model inspiring several works in the literature is proposed by Jadbabaie et al. \cite{jadbabaie2012non}. In this model, a fixed true {\it state} or {\it hypothesis} is aimed to be recovered by a network of agents. The state might be a decision or opinion of an individual, the correct price of a product or any quantity of interest. The state is assumed to belong to a {\it finite} collection of states. Agents receive a stream of {\it private} signals about the true state. However, the signals do {\it not} provide enough information for agents to detect the underlying state of the world. Hence, agents engage in {\it local} interactions to compensate for their imperfect knowledge about the states. Numerous works build on this approach to study social learning and distributed detection \cite{shahrampour2013exponentially,lalitha2014social,shahrampour2015switching}. The focus of theses works is the asymptotic behavior of the model. Though appealing in certain cases, asymptotic analysis might not unveil all important parameters for learning. Therefore, {\it finite-time} analysis of the problem is also an interesting complementary direction to study.

Let us first provide more details on the asymptotic analysis of the problem. The authors in \cite{jadbabaie2012non} propose a non-Bayesian update rule for social learning applications. Each individual updates her Bayesian prior, and averages the result with the opinion of her neighbors. Under mild technical assumptions, the authors prove that agents' beliefs converge to a delta distribution on the true state. The convergence occurs exponentially fast and in almost sure sense. Shahrampour and Jadbabaie \cite{shahrampour2013exponentially} consider an optimization-based approach to the problem, inspired by the work of Duchi et al. \cite{duchi2012dual} on distributed dual averaging. Their proposed update rule is essentially a distributed stochastic optimization. They establish that the sequence of beliefs is weakly consistent (convergence in probability) when agents employ a {\it gossip} communication protocol.  A communication-efficient variant of the problem is considered in \cite{shahrampour2015switching} in which agents switch between Bayesian and non-Bayesian regimes to detect the true state. Furthermore, Lalitha et al. \cite{lalitha2014social} develop a strategy where agents perform a Bayesian update, and geometrically average the posteriors in their local neighborhood. The authors then provide the convergence and rate analysis of their method. In \cite{jadbabaie2012non,shahrampour2013exponentially,shahrampour2015switching,lalitha2014social}, the convergence occurs exponentially fast, and the {\it asymptotic} rate is characterized via the {\it relative entropy} of individuals' signal structures and their {\it eigenvector centralities} (in directed networks).

The asymptotic analysis presented in the works above only characterizes the dominant factors in the long run. In real world applications, however, the {\it decision} has to be made in a {\it finite} time horizon. Hence, it is important to study the finite-time version of the problem to understand the role of {\it network parameters} in detection quality. Serving this goal, the works of \cite{shahrampour2014distributed,7172262,nedic2015fast} study the {\it non-asymptotic} problem. While the network structure in \cite{shahrampour2014distributed} is assumed to be fixed, the works of \cite{7172262,nedic2015fast} address switching protocols that are deterministic.

In this paper we build on the results of \cite{shahrampour2014distributed} to extend our setup to random networks. For fixed networks, we define the notion of {\it Kullback-Leibler (KL)} cost to compare the performance of distributed setting to its centralized counterpart. We provide an upper bound on the cost in terms of the {\it spectral gap} of the network. Our cost bound is independent of time with {\it high probability}. We further consider the stochastic communication setting in which the structure is realized randomly at each iteration. We prove that in this case, the rate scales inversely in the spectral gap of the expected network. Our result also guarantees the almost sure learning in random networks.

The rest of the paper is organized as follows: we formalize the notation, problem and the distributed detection scheme in Section \ref{The Problem Description and Algorithm}. Section \ref{Finite-time Analysis} is devoted to the finite-time analysis of the problem. We consider both fixed and switching network topologies and provide non-asymptotic results. Section \ref{Conclusion} provides the concluding remarks, and proofs are included in the Appendix.

\section{The Problem Description and Algorithm}\label{The Problem Description and Algorithm}
\subsection{Notation} 
We use the following notation throughout the paper: {\small
\begin{center}
  \begin{tabular}{| c || l | }
    \hline
     $[n]$ &  The set $\{1,2,...,n\}$ for any integer $n$ \\ \hline
     $x^\tr$ & Transpose of the vector $x$ \\ \hline
     $x(k)$ & The $k$-th element of vector $x$ \\ \hline
     $\base_k $ & Delta distribution on $k$-th component \\ \hline
      $\mathbb{1}$ & Vector of all ones \\ \hline
     $\|\mu-\pi\|_{\text{TV}}$ &  Total variation distance between $\mu,\pi \in \Delta_m$ \\ \hline
     $D_{KL}(\mu \| \pi)$ &  KL-divergence of $\pi\in \Delta_m$ from $\mu \in \Delta_m$  \\ \hline
     $\sigma_i(W)$ &  The $i$-th largest singular value of matrix $W$ \\ \hline
  \end{tabular}
\end{center}}
Furthermore, all the vectors are assumed to be in column format.

\subsection{Observation Model}
We consider a setting in which $\Theta=\{\theta_1,\theta_2,\ldots,\theta_m\}$ denotes a finite set of {\it states} of the world. A network of $n$ agents seek the {\it unique}, true state of the world $\theta_1\in \Theta$ (unknown of the problem). At each time $t\in [T]$, the belief of agent $i$ is represented by  $\mu_{i,t} \in  \Delta_m$, where $\Delta_m$ is a probability distribution over the set $\Theta$. For instance, $\mu_{i,0} \in \Delta_m$ denotes the prior belief of agent $i \in [n]$ about the states of the world, and it is assumed to be uniform\footnote{The assumption of uniform prior only avoids notational clutter. The analysis in the paper holds for any prior with full support.}. 

The detection model is defined with a conditional likelihood function $\ell(\cdot|\theta_k)$ parametrized by some state of the world $\theta_k \in \Theta$. For each $i\in [n]$, $\ell_i(\cdot|\theta_k)$ denotes the $i$-th marginal of $\ell(\cdot|\theta_k)$, and we use the vector representation $\ell_i(\cdot|\theta)=[\ell_i(\cdot|\theta_1),...,\ell_i(\cdot|\theta_m)]^\tr$ to stack all states. At each time $t\in [T]$, the signal $s_t=(s_{1,t},s_{2,t},\ldots,s_{n,t})\in \mathcal{S}_1\times \dots \times \mathcal{S}_n$ is generated based on the true state $\theta_1$. As a result, for each $i \in [n]$, the signal $s_{i,t}\in \mathcal{S}_i$ is a sample drawn from the likelihood $\ell(\cdot|\theta_1)$ where $\mathcal{S}_i$ is the sample space. 

The signals are i.i.d. over time horizon, and also the marginals are independent across agents, i.e., $\ell(\cdot|\theta_k)=\Pi_{i=1}^n \ell_i(\cdot|\theta_k)$ for any $k \in [m]$. For simplicity, we define the log-marginal $\psi_{i,t}:= \log \ell_i(s_{i,t} |\theta)$ which is a sample corresponding to $\Psi_i:= \log \ell_i(\cdot|\theta)$ for any $i\in [n]$.
\begin{description}
\item[{\bf A1.}] We assume that all log-marginals are uniformly bounded such that $\|\psi_{i,t}\|_\infty \leq B$ for any $s_{i,t}\in \mathcal{S}_i$, i.e., we have $|\log \ell_i(\cdot |\theta_k)| \leq B$ for any $i \in [n]$ and $k\in [m]$.
\end{description}
Based on assumption {\bf A1}, every private signal has bounded information content \cite{drakopoulos2013learning}. This bound can be found, for instance, when the signal space is discrete and provides a full support for distribution. Let us define $\bar{\Theta}_i$ as the set of states that are observationally equivalent to $\theta_1$ for agent $i\in [n]$; in other words, $\bar{\Theta}_i=\{\theta_k \in \Theta : \ell_i(s_i|\theta_k)= \ell_i(s_i| \theta_1) \ \ \forall s_i\in \mathcal{S}_i\} $ almost surely with respect to the signal space. As evident from the definition, any state $\theta_k \neq \theta_1$ in the set $ \bar{\Theta}_i$ is not distinguishable from the true state by observation of samples from the $i$-th marginal. Let $\bar{\Theta}=\cap_{i=1}^n\bar{\Theta}_i$ be the set of states that are observationally equivalent to $\theta_1$ from all agents perspective. 
\begin{description}
\item[{\bf A2.}] We assume that no state in the world is observationally equivalent to the true state from the standpoint of the network, i.e., the true state is globally identifiable, and we have $\bar{\Theta}=\{\theta_1\}$. 
\end{description}
Assumption {\bf A2} guarantees that agents can benefit from collaboration with each other. In other words, for any false state $\theta_k \neq \theta_1$, there must exist an agent who is able to distinguish $\theta_1$ from $\theta_k$.


Finally, the operator $\Ex_{\mathcal{S}}[\cdot]$ denotes the expectation with respect to signal space throughout the paper.  

\subsection{Network Model}\label{Network Model}
Private signals are not informative enough for agents, so they interact with each other to learn the true state of the world. For any time $t \in [T]$, the graph $G_t=([n],E_t)$ captures the network structure, i.e. $[n]$ is the set of nodes corresponding to agents ($n>1$), and $E_t$ is the set of edges for that particular round. Agent $i$ receives information from $j$ only if  the pair $(i,j) \in E_t$. We also define the neighborhood of agent $i$  at any time $t \in [T]$ as $\mathcal{N}_{i,t}:=\left\{j\in [n]: (i,j)\in E_t\right\}$. We represent by $[W(t)]_{ii}> 0$ the {\it self-reliance} of agent $i$ at time $t$, and by $[W(t)]_{ij}>0$ the weight that agent $i$ assigns to information received from agent $j$ in round $t$. The matrix $W(t)$ is then defined such that $[W(t)]_{ij}$ denotes the entry in its $i$-th row and $j$-th column. By construction, $W(t)$ has nonnegative entries, and $[W(t)]_{ij}>0$ only if $(i,j)\in E_t$. We further assume that $W(t)$ is doubly stochastic and symmetric; hence,
\begin{align*}
W(t)^\top=W(t) \ \ \ \  \ \sum_{j=1}^n[W(t)]_{ij}=\sum_{j\in \mathcal{N}_{i,t}}[W(t)]_{ij}=1.
\end{align*}
To form the network structure, $W(t)$ is drawn {\it independently} over time from a {\it stationary} distribution $\mathcal{W}$, i.e. the elements of the sequence $\{W(t)\}_{t=1}^T$ are i.i.d. samples. Note importantly that the source of randomness is independent of signal space. We distinguish any expectation with respect to the network randomness from signal space using $\Ex_{\mathcal{W}}[\cdot]$.
\begin{description}
\item[{\bf A3.}] The network is \textit{connected} in expectation sense. That is, for matrix $W=\Ex_{\mathcal{W}}[W(t)]$ there exists a bidirectional path from any agent $i\in [n]$ to any agent $j\in [n]$. 
 \end{description}
 The assumption guarantees that (in expectation sense) information can be propagated properly throughout the network. It further implies that $\sigma_1(W)=1$ is unique, and the other singular values  of $W$ are strictly less than one in magnitude \cite{rosenthal1995convergence}. Since the matrix $W$ is symmetric, each agent in the network is {\it equally} central, and 
\begin{align*}
\mathbb{1}^\top W = \mathbb{1}.
\end{align*}
Therefore, $\mathbb{1}/n$ is the stationary distribution of $W$. Assumption {\bf A3} entails that the Markov chain $W$ is irreducible and aperiodic \cite{rosenthal1995convergence}. 
 
A well-known example of the communication protocol detailed above is {\it gossip} algorithm. The communication protocol works based on a Poisson clock. Once the clock ticks, an agent from the set $[n]$ is picked uniformly at random. The agent then selects a neighboring agent (with respect to a fixed, predefined structure) at random, and they average their information. For a thorough review of gossip algorithms we refer the interested reader to \cite{boyd2006randomized}.   

\subsection{Centralized vs. Decentralized Detection}
Decentralized detection is constructed based on its centralized analog. In the centralized scenario there is only one agent in the world (no network exists), and the agent has {\it global} information to learn the true state. In other words, the agent has full access to the sequence of signals $\{s_t\}_{t=1}^T$. At any round $t \in [T]$, the agent accumulates an {\it empirical average} of log-marginals, and forms the {\it belief} $\mu_t \in \Delta_m$ about the states, where $\Delta_m=\{\mu \in \Real^m \ | \ \mu \succeq 0, \ \sum_{k=1}^m\mu(k)=1\}$ denotes the $m$-dimensional probability simplex.  
Defining 
\begin{align}\label{psi definition} 
\psi_{t}:=  \frac{1}{n}\sum_{i=1}^n \psi_{i,t}=\frac{1}{n}\sum_{i=1}^n \log \ell_i(s_{i,t}|\theta),
\end{align}
the following updates capture the centralized detection
\begin{align}
\phi_{t}&=\phi_{t-1}+ \psi_{t} \ \ \  , \ \ \ \mu_t(k)=\frac{\exp\{\eta \phi_t(k)\}}{\sum_{z=1}^m\exp\{\eta \phi_t(z)\}} \label{CenBelief}.
\end{align}
It can be seen from above that the centralized detector aggregates a geometric average of marginals in $\phi_t$. The parameter $\eta$ is called the {\it learning rate}.

One can prove that (see e.g. \cite{shahrampour2014distributed}) the following inequality holds 
\begin{align*}
\Ex\left[\sum_{i=1}^n \Psi_i(k)\right]&=\Ex\left[\sum_{i=1}^n  \log \ell_i(\cdot  |\theta_k)\right] \\
&< \Ex\left[\sum_{i=1}^n  \log \ell_i(\cdot  |\theta_1)\right]=\Ex\left[\sum_{i=1}^n  \Psi_i(1)\right],
\end{align*}
for any $k\neq 1$, due to uniqueness of the true state $\theta_1$ (assumption {\bf A2}). In what follows, without loss of generality, we assume the following descending order, i.e. 
  \begin{align*}
\Ex\left[\sum_{i=1}^n  \Psi_i(1)\right]&>\Ex\left[\sum_{i=1}^n  \Psi_i(2)\right]\geq \cdots \geq \Ex\left[\sum_{i=1}^n  \Psi_i(m)\right], \label{orderassum}\numberthis
\end{align*}
The assumption will only help us to simplify the derivation of technical results.

We now describe the distributed setting which involves a network of agents. In this scenario, each agent $i\in [n]$ only receives a stream of private signals $\{s_{i,t}\}_{t=1}^T$ generated based on the parametrized likelihood $\ell_i(\cdot|\theta_1)$. Therefore, agent $i\in [n]$ does not directly observe $s_{j,t}$ for any $j\neq i$. In other words, the agent collects local information by calculating a weighted average of log-likelihoods in its neighborhood. Then, the agent forms the belief $\mu_{i,t}\in \Delta_m$ at round $t\in [T]$ as follows:
\begin{align}\label{DecBelief}
\phi_{i,t}&=\sum_{j \in \mathcal{N}_{i,t}} [W(t)]_{ij}\phi_{j,t-1}+ \psi_{i,t} \notag \\ 
\mu_{i,t}(k)&=\frac{\exp\{\eta \phi_{i,t}(k)\}}{\sum_{z=1}^m\exp\{\eta \phi_{i,t}(z)\}}.
\end{align}

As depicted above, each agent updates its belief using purely local diffusion. Let us distinguish the centralized and decentralized detector more specifically. The centralized detector collects {\it all} log-marginals, whereas the decentralized detector receives {\it private} log-marginals, and collects a weighted average of {\it local} information. Regardless of the information collection part, both algorithms are special cases of the well-known {\it Exponential Weights} algorithm. 

It can be verified (see e.g. \cite{shahrampour2013exponentially}) that the closed-form solution of $\phi_{i,t}$ in the decentralized update \eqref{DecBelief} is as follows 
\begin{align*}
\phi_{i,t}= \sum_{\tau=1}^{t}\sum_{j=1}^n\left[\prod_{\rho=0}^{t-1-\tau}W(t-\rho)\right]_{ij}\psi_{j,\tau},
\end{align*}
for any $i \in [n]$. One can also combine above with \eqref{psi definition} and \eqref{CenBelief} to observe that
\begin{align*}
\frac{1}{n}\sum_{i=1}^n\phi_{i,t}&=\frac{1}{n}\sum_{\tau=1}^{t}\sum_{j=1}^n\sum_{i=1}^n\left[\prod_{\rho=0}^{t-1-\tau}W(t-\rho)\right]_{ij}\psi_{j,\tau}\\
&=\frac{1}{n}\sum_{\tau=1}^{t}\sum_{j=1}^n\psi_{j,\tau}=\phi_t,
\end{align*}
since product of doubly stochastic matrices remains doubly stochastic. The identity above draws the connection between centralized and decentralized updates.

\section{Finite-time Analysis}\label{Finite-time Analysis}
In this section, we provide our technical results. We first specialize the problem to {\it fixed} network structures, and then consider {\it switching} topologies, and prove non-asymptotic results for convergence of beliefs.

\subsection{Fixed Network Topology}
One can investigate the convergence of beliefs in fixed networks; however, we present a more general result in this section. We measure the efficiency of the distributed algorithm versus its centralized counterpart using the notion of {\it decentralization} cost. Throughout this section we assume that $W(t)=W$ with probability one for all $t\in [T]$.

At any round $t\in [T]$ , we quantify the cost which agent $i\in [n]$ needs to pay to have the same opinion as a centralized agent with $D_{KL}(\mu_{i,t}\| \mu_t)$; then, the agent suffers a total decentralization cost of
\begin{align}\label{Regret}
\R_{i,T}:= \sum_{t=1}^T D_{KL}(\mu_{i,t}\| \mu_t),
\end{align}
after $T$ rounds. In general, the KL-divergence measures the dissimilarity of two probability distributions; hence, it could be a reasonable metric to capture the difference between two algorithms as they both output a probability distribution over state space. 

More formally, the cost quantifies the difference between a decentralized agent that observes private signals $\{s_{i,t}\}_{t=1}^T$ and a centralized agent that has $\{s_t\}_{t=1}^T$ available. In other words, it shows how well the decentralized algorithm copes with the partial information. It is important to note that $\R_{i,T}$ is still a random quantity as it depends on signals. Our goal is to find a bound on the cost in the {\it high probability} sense.

The connectivity of network plays an important role in the learning as $W^t \rightarrow \frac{1}{n}\mathbb{1}\mathbb{1}^\tr$ as $t\rightarrow \infty$. We shall see that the cost bound is governed by the mixture behavior of Markov chain $W$. We now present the main result of this section in the following theorem. The proof can be found in \cite{shahrampour2014distributed}.
\begin{theorem}\label{DecRegretRate}
Let the sequence of beliefs $\{\mu_{i,t}\}_{t=1}^T$ for each agent $i\in [n]$ be generated by the Distributed Detection algorithm with the choice of learning rate $\eta=\frac{1-\sigma_2(W)}{16B\log n}$. Given bounded log-marginals (assumption {\bf A1}), global identifiability of the true state (assumption {\bf A2}), and connectivity of the network (assumption {\bf A3} with $W(t)=W$ a.e.), we have 
\begin{align*}
 \R_{i,T} &\leq \frac{18 B^2}{{\mathcal{I}^2(\theta_1,\theta_2)}} \max \left\{  \log\frac{6m}{\delta} , \frac{3B\sqrt{2}}{\mathcal{I}(\theta_1,\theta_2)}  \right\}\\
 &+ \frac{48B\log n}{\mathcal{I}(\theta_1,\theta_2)}\frac{\log m+2}{1-\lambda_{\max}(W)},
\end{align*}
with probability at least $1-\delta$. 
\end{theorem}
We remark that the special choice of learning rate only simplifies the bound. We do not tune the learning rate for optimization purposes; otherwise, using the same $\eta$ for both algorithms would not provide a fair comparison. One can also work with $\eta=1$ for both algorithms, and derive a bound which looks slightly more complicated than the bound in Theorem \ref{DecRegretRate}.

The dependence of bound to the inverse of $\mathcal{I}(\theta_1,\theta_2)$ is quite natural since it can be seen as the asymptotic convergence rate of beliefs (see e.g. \cite{shahrampour2013exponentially,shahrampour2015switching,lalitha2014social}). It simply means that when observations under $\theta_2$ (the second likeliest state) are as likely as observations under $\theta_1$ (the true state), the cost of the algorithm increases. Intuitively, when signals hardly reveal the difference between the best two candidates for the true state, agents must spend more effort to discern the two. Hence, this results in suffering a larger cost caused by slower learning rate. 

The cost scales logarithmically with the network size $n$ and the number of states $m$. It further scales inversely in 
\begin{align*}
\gamma(W):= 1-\sigma_2(W), 
\end{align*}
defined as the {\it spectral gap} of the network. Interestingly, in a fixed network, the detection cost is time-independent (with high probability), showing the best behavior with respect to time. Therefore, the average expected cost (per iteration cost) asymptotically tends to zero. We finally indicate that dependence of the bound to $\sigma_2(W)$ is important from network design perspective \cite{shahrampour2014distributed}.

\subsection{Switching Network Topology}
In this section, we investigate the convergence of agents' beliefs in {\it time-varying} networks. As described in Section \ref{Network Model}, at every time $t$ the network structure is realized with a random matrix $W(t)$. Agents then communicate with each other following $W(t)$. We assume that the matrix is generated from a {\it stationary} distribution, and therefore, we have $\Ex_{\mathcal{W}}[W(t)]=W$ for all $t\in [T]$. Assumption {\bf A3} guarantees that the network is connected in {\it expectation} sense. We show that even in {\it time-varying} networks the mixture behavior of $W$ (expected network) affects convergence of beliefs with high probability.

We now establish that agents have arbitrarily close opinions in a network that is connected in expectation sense. The following proposition proves that the convergence rate is governed by relative entropy of signals and network characteristics. 
\begin{proposition}\label{Distributed Beliefs}
Let the sequence of beliefs $\{\mu_{i,t}\}_{t=1}^T$ for each agent $i\in [n]$ be generated by the Distributed Detection algorithm \eqref{DecBelief} with the learning rate $\eta=1$. Given bounded log-marginals (assumption {\bf A1}), global identifiability of the true state (assumption {\bf A2}), and connectivity of the expected network (assumption {\bf A3}), for each individual agent $i\in [n]$ it holds that
\begin{align*}
\log \|\mu_{i,t}-\base_1\|_{\text{TV}} &\leq - \mathcal{I}(\theta_1,\theta_2) t  + \sqrt{2B^2t\log \frac{m}{\delta}} \\
&+ \frac{8B\log n}{1-\sigma_2(W)}+\log m,
\end{align*}
with probability at least $1-\delta$, where for $k\geq 2$
\begin{align*}
\mathcal{I}(\theta_1,\theta_k):= \frac{1}{n}\sum_{i=1}^nD_{KL}(\ell_i(\cdot|\theta_1)\|\ell_i(\cdot|\theta_k)).
\end{align*}
\end{proposition}
The proposition provides an {\it anytime} bound on the log-distance in the high probability sense. It also verifies that the belief $\mu_{i,t}$ of each agent $i\in [n]$ is {\it strongly} consistent, i.e., it converges almost surely to a delta distribution on the true state. It is evident that the dominant term (asymptotic rate) depends on relative entropy of signals through $\mathcal{I}(\theta_1,\theta_2)$. This is consistent with previous asymptotic results found in \cite{shahrampour2013exponentially,shahrampour2015switching,lalitha2014social} for other updates similar to our update. Proposition \ref{Distributed Beliefs} complements those results by providing a non-asymptotic convergence rate. Finally, the inverse scaling with spectral gap also remains in effect (for expected network) similar to the case of fixed topologies.



\section{Conclusion}\label{Conclusion}
We considered the distributed detection problem in fixed and switching network topologies. A network of agents observe a stream of private signals which do not provide enough information about the true state of the world. Therefore, agents must communicate with each other to augment their imperfect knowledge with local observations. Iteratively forming a belief about the states, each agent uses purely local diffusion to update itself. We analyzed the detection problem in {\it finite-time} domain. We first specialized to the case of fixed networks, and study the efficiency of our algorithm versus its centralized analog. We introduced a KL cost to measure the dissimilarity of the two algorithms, and bounded the cost in terms of relative entropy of signals, network size and spectral gap. We further extended our results to {\it switching} network topologies. We investigated convergence of beliefs, and provided an anytime bound on the detection error in the high probability sense. In this case, the spectral gap of the expected network proves to be crucial in convergence rate. As a future direction, we would like to consider the scenario where the signal distribution is not stationary. Studying drifting distributions allows us to examine the robustness of detection in dynamic framework.  

\section*{Appendix : Omitted Proofs}
We use McDiarmid's inequality in Lemma \ref{Mc} for the proof of Proposition \ref{Distributed Beliefs}.

\noindent
\textbf{\emph{Proof of Proposition \ref{Distributed Beliefs}}}. Letting $\eta=1$ in \eqref{DecBelief}, we write
\begin{align*}
\mu_{i,t}(1)&=\frac{\exp\left\{ \phi_{i,t}(1)\right\}}{\sum_{k=1}^m\exp\left\{ \phi_{i,t}(k)\right\}}\\
&=\left(1+\sum_{k=2}^m\exp\left\{ \phi_{i,t}(k)- \phi_{i,t}(1)\right\}\right)^{-1}\\
&\geq 1-\sum_{k=2}^m\exp\left\{ \phi_{i,t}(k)-\phi_{i,t}(1)\right\}, \label{E1}\numberthis
\end{align*}
using the simple inequality that $(1+x)^{-1}\geq 1-x$ for any $x \geq 0$. Since we know
\begin{align*}
\|\mu_{i,t}-\base_1\|_{\text{TV}}&=\frac{1}{2}\left(1-\mu_{i,t}(1)+\sum_{k=2}^m\mu_{i,t}(k)\right)\\
&=1-\mu_{i,t}(1),
\end{align*}
we can combine above with \eqref{E1} to get
\begin{align}
\|\mu_{i,t}-\base_1\|_{\text{TV}} \leq \sum_{k=2}^m\exp\left\{ \phi_{i,t}(k)- \phi_{i,t}(1)\right\}.\label{E2}
\end{align} 
For any $k\in [m]$, define
\begin{align*}
\Phi_{i,t}(k):= \sum_{\tau=1}^{t}\sum_{j=1}^n\left[\prod_{\rho=0}^{t-1-\tau}W(t-\rho)\right]_{ij}\log \ell_j(\cdot |\theta_k),
\end{align*}
and note that given $\{W(\tau)\}_{\tau=1}^t$, $\Phi_{i,t}(k)$ is a function of $nt$ random variables $\{s_{i,\tau}\}_{\tau=1}^t$ for all $i\in [n]$. To use McDiarmid's inequality in Lemma \ref{Mc}, set $H=\Phi_{i,t}(k)-\Phi_{i,t}(1)$, fix the samples for $nt-1$ random variables, and draw two different samples $s_{j,\tau}$ and $s'_{j,\tau}$ for some $j\in [n]$ and some $\tau \in [t]$. The fixed samples are simply cancelled in the subtraction, and we derive

\vspace{-0.22in}
{\small
\begin{align*}
&|H(...,s_{j,\tau},...)-H(...,s'_{j,\tau},...)|\\
&~~~~~=\left[\prod_{\rho=0}^{t-1-\tau}W(t-\rho)\right]_{ij}\left|\log \frac{\ell_j(s_{j,\tau} |\theta_k)}{\ell_j(s_{j,\tau} |\theta_1)}-\log\frac{\ell_j(s'_{j,\tau} |\theta_k)}{\ell_j(s'_{j,\tau} |\theta_1)}\right|\\
&~~~~~\leq \left[\prod_{\rho=0}^{t-1-\tau}W(t-\rho)\right]_{ij} 2B,
\end{align*}}
where we applied assumption {\bf A1}. Since the product of doubly stochastic matrices remains doubly stochastic, summing over $j\in [n]$ and $\tau \in [t]$, we obtain
\begin{align*}
\sum_{\tau=1}^{t}\sum_{j=1}^n\left(\left[\prod_{\rho=0}^{t-1-\tau}W(t-\rho)\right]_{ij} 2B\right)^2 \leq 4B^2t.
\end{align*}
Let us define the event $A$ as
\begin{align*}
A:=\Big\{\phi_{i,t}(k)-\phi_{i,t}(1)>\Ex\left[\Phi_{i,t}(k)-\Phi_{i,t}(1)\right]+\varepsilon \Big\},
\end{align*}
where $\Ex[\cdot]$ is expectation over all sources of randomness (signals and network structures). We then have
\begin{align*}
\Prob(A)=\Ex_{\mathcal{W}}\left[\Prob\big(A \big| \{W(\tau)\}_{\tau=1}^t \big)\right],
\end{align*}
where the expectation $\Ex_{\mathcal{W}}[\cdot]$ is taken with respect to network randomness. We now apply McDiarmid's inequality in Lemma \ref{Mc} to obtain
\begin{align*}
\Prob\left(A\big| \{W(\tau)\}_{\tau=1}^t \right) \leq \exp\left\{\frac{-\varepsilon^2}{2B^2t}\right\},
\end{align*}
for each fixed $k$. Letting the probability above equal to $\delta/m$ and taking a union bound over $k\in [m]$, the following event $(A^C)$ holds (given $\{W(\tau)\}_{\tau=1}^t$)
\begin{align}
\phi_{i,t}(k)-\phi_{i,t}(1)\leq \Ex\left[\Phi_{i,t}(k)-\Phi_{i,t}(1)\right]+\sqrt{2B^2t\log \frac{m}{\delta} }, \label{E7}
\end{align}
simultaneously for all $k=2,...,m$, with probability at least $1-\delta$, which implies
\begin{align*}
\Prob(A^C)=\Ex_{\mathcal{W}}\left[\Prob\big(A^C \big| \{W(\tau)\}_{\tau=1}^t \big)\right] \geq 1-\delta.
\end{align*}
On the other hand, in view of assumption {\bf A1} and the fact that $\Ex_{\mathcal{W}}[W(t)]=W$ for all $t$, we have
\begin{align*}
&\Ex\left[\Phi_{i,t}(k)-\Phi_{i,t}(1)\right] \\
&~~~=\sum_{\tau=1}^{t}\sum_{j=1}^n\left[W^{t-\tau}\right]_{ij}\Ex_{\mathcal{S}}\left[\log \ell_j(\cdot |\theta_k)-\log \ell_j(\cdot |\theta_1)\right]\\
&~~~=\sum_{\tau=1}^{t}\sum_{j=1}^n\left(\left[W^{t-\tau}\right]_{ij}-\frac{1}{n}\right)\Ex_{\mathcal{S}}\left[\log \ell_j(\cdot |\theta_k)-\log \ell_j(\cdot |\theta_1)\right]\\
&~~~+\frac{1}{n}\sum_{\tau=1}^{t}\sum_{j=1}^n \Ex_{\mathcal{S}}\left[\log \ell_j(\cdot |\theta_k)-\log \ell_j(\cdot |\theta_1)\right]\\
&~~~\leq 2B\sum_{\tau=1}^{t}\sum_{j=1}^n\left|\left[W^{t-\tau}\right]_{ij}-\frac{1}{n}\right|\\
&~~~-\frac{t}{n}\sum_{j=1}^nD_{KL}\left(\ell_j(\cdot |\theta_1)\| \ell_j(\cdot |\theta_k)\right)\\
&~~~= 2B\sum_{\tau=1}^{t}\sum_{j=1}^n\left|\left[W^{t-\tau}\right]_{ij}-\frac{1}{n}\right|-\mathcal{I}(\theta_1,\theta_k)t \\
&~~~\leq \frac{8B\log n}{1-\sigma_2(W)}-\mathcal{I}(\theta_1,\theta_k)t,
\end{align*}
where we applied Lemma 2 in \cite{shahrampour2014distributed} to derive the last step. Using \eqref{orderassum}, we simplify above to get 
\begin{align}
\Ex\left[\Phi_{i,t}(k)-\Phi_{i,t}(1)\right]&\leq \frac{8B\log n}{1-\sigma_2(W)}-\mathcal{I}(\theta_1,\theta_2)t, \label{E5}
\end{align}
for any $k=2,...,m$. Substituting \eqref{E5} into \eqref{E7} and combining with \eqref{E2}, we have

\vspace{-0.15in}
{\small
\begin{align*}
&\|\mu_{i,t}-\base_1\|_{\text{TV}} \\
&~~\leq \sum_{k=2}^m \exp \left\{-\mathcal{I}(\theta_1,\theta_2)t+ \sqrt{2B^2t\log \frac{m}{\delta} }+\frac{8 B\log n}{1-\sigma_2(W)}\right\}\\
&~~\leq m \exp \left\{-\mathcal{I}(\theta_1,\theta_2)t+ \sqrt{2B^2t\log \frac{m}{\delta} }+\frac{8 B\log n}{1-\sigma_2(W)}\right\},
\end{align*} }with probability at least $1-\delta$, and the proof is completed.$\hfill \blacksquare $\\

\begin{lemma}{\bf (McDiarmid's Inequality)}\label{Mc} 
Let $X_1,...,X_N \in \chi$ be independent random variables and consider the mapping $H: \chi^N \mapsto \Real$. If for $i\in \{1,...,N\}$, and every sample $x_1,...,x_N,x'_i\in \chi$, the function $H$ satisfies
\begin{align*}
\left|H(x_1,...,x_{i},...,x_N)-H(x_1,...,x'_{i},...,x_N)\right| \leq c_i,
\end{align*}
then for all $\varepsilon>0$, {\small
\begin{align*}
\Prob\bigg\{H(x_1,...,x_N)-\Ex\left[H(X_1,...,X_N)\right] \geq \varepsilon\bigg\}\leq \exp\left\{\frac{-2\varepsilon^2}{\sum_{i=1}^Nc_i^2}\right\}.
\end{align*}}
\end{lemma}

\bibliographystyle{IEEEtran}
\bibliography{IEEEabrv,shahin}

\end{document}